\documentclass[11pt,graphicx,amsmath]{article}
\usepackage{amsmath}
\usepackage{graphicx}
\usepackage{bm}
\usepackage{color}
\usepackage{amssymb}
\usepackage{amsfonts}
\usepackage{comment}
\usepackage{cite}
\usepackage{todonotes}
\usepackage{caption}
\usepackage{subcaption}
\usepackage{appendix}
\usepackage{epstopdf}

\def\be{\begin{equation}}
\def\ee{\end{equation}}
\def\nn{\nonumber}

\def\ba{\begin{eqnarray}}
\def\ea{\end{eqnarray}}
\def\bl#1\el{\begin{align}#1\end{align}}

\def\be{\begin{equation}}
\def\ee{\end{equation}}
\def\ba{\begin{eqnarray}}
\def\ea{\end{eqnarray}}
\def\nn{\nonumber}
\def\bl#1\el{\begin{align}#1\end{align}}

\title{  The evolution of two-point correlation function of galaxies
          with a twin-peak initial power spectrum }
\author{\small
            \,  Yang  Zhang  \thanks{yzh@ustc.edu.cn} , \,
             Bichu Li  \thanks{libichu@mail.ustc.edu.cn}     \\
 \small  Department of  Astronomy,
   CAS Key Laboratory for Researches in Galaxies and Cosmology, \\
 \small  University of Science and Technology of China, Hefei, Anhui, 230026, China \\
 }

 \date{}

\begin{document}

\maketitle

\begin{abstract}


The evolution equation of two-point correlation function $\xi$ of galaxies
can analytically describe the large scale structure of the galaxy distribution,
and the solution depends also upon the initial condition.
The primeval spectrum of the baryon acoustic oscillations (BAO)
contains multi peaks that survived the Silk damping,
and,  as a relevant portion, two peaks of the  primeval BAO spectrum
fall into the range of current galaxy surveys.
Incorporating this portion,
we use a twin-peak initial  power spectrum of the galaxies,
and obtain the evolution solution from a redshift $z=8$ to $z=0$
 in the Gaussian approximation.
The outcome $\xi(r)$ at $z=0.6$ still exhibits the 100 Mpc periodic bumps
as observed  by  the WiggleZ survey,
a feature largely determined by the Jeans length $\lambda_J$ in the equation.
In particular, due to the superposition of the twin peaks in the initial condition,
$\xi(r)$ shows a shallow trough at $\sim 70 h^{-1}$Mpc
and a deep trough at $\sim 140 h^{-1}$Mpc,
agreeing with the observational data,
much better than our previous work that used a simple one-peak initial spectrum.

\end{abstract}

\

          and classical fields in curved spacetime

98.65.Dx  Large-scale structure of the Universe

98.80.Jk Mathematical  aspects of cosmology

\newpage

\section{Introduction }

Progress in the observations of
large-scale structure is ever increasing.
Surveys for galaxies, e.g., 6dFGS \cite{Beutler6dFGS2011},
SDSS \cite{Anderson2012,Anderson2014,SanchezSDSS2017,Alam2017,Ross2015,Bautista2021,Gil-Marin2020},
WiggleZ \cite{RuggeriBlake2020},
and for quasars,
e.g., SDSS BOSS \cite{Busca2013,Slosar2013,BausticaBusca2017,Ata2018,Agathe2019,Neveux2020},
have provide rich data  of the distribution of galaxies.
Most of theoretical studies are  numerical computations and simulations
since late 70's.
Attempts on analytic study are also made in various approach
\cite{Saslaw1985,deVega1996}.
In particular, Davis and Peebles \cite{DaviesPeebles1977}
treated the system of galaxies as a many-body system,
started  with the Liouville's equation of probability function,
and derived a set of BBGKY (Bogoliubov-Born-Green-Kirkwood-Yvon)
equations of the  two-point correlation function $\xi(r,t)$
and of the velocity dispersions of galaxies.
In particular, the equation of $\xi(r,t)$  is not closed,
and contains other   unknowns.
To solve  these five equations,
an appropriate initial condition is required for the five unknown functions,
which are difficult to specify consistently.
Due to these points,  the BBGKY equations are hard to
apply in the practical study  of the large scale structure.

In our serial analytic study,
using the field theoretic technique,
the static equations and their solutions have been given
for the two-point correlation functions  \cite{Zhang2007,ZhangMiao2009,ZhangChen2015,ZhangChenWu2019},
for the three-point correlation functions \cite{ZhangChenWu2019,WuZhang2022-2},
and for the four-point correlation functions \cite{ZhangWu2024}.
To describe the evolution in the expanding Universe,
in Ref.\cite{2021ZhangLi},
we derived the evolution equation of two-point correlation function
in a flat Robertson-Walker spacetime,
which is a closed, nonlinear, differential-integro equation.
Using a simple one-peak initial power spectrum,
we solved the evolution equation in the Gaussian approximation,
and the solution $\xi(r,t)$ showed the 100Mpc periodic bumps
as have been observed by the early surveys
\cite{Broadhurst1990,Broadhurst1995,Tucker1997,Einasto1997b,Einasto2002b,Tago2002},
and by more recent surveys of galaxy
\cite{RuggeriBlake2020,Beutler6dFGS2011,Anderson2014,Anderson2012,SanchezSDSS2017},
and of quasars \cite{Ata2018,BausticaBusca2017,Busca2013,Slosar2013,Agathe2019,Neveux2020}.
However, the depths of the predicted two troughs at
$70 h^{-1}$Mpc and $140 h^{-1}$Mpc,
still deviate from the observational data.
In this paper, we shall use the twin-peak initial power spectrum
that inherits a relevant portion of
the primeval BAO spectrum
\cite{SunyaevZeldovich1970,PeeblesYu1970,EisensteinHu1998,HuSugiyama1995,Holtzman1989}.
The solution will not only yield the 100Mpc periodic feature,
but also predict the adequate depths of the two troughs,
which agree with the observational data of the WiggleZ Survey \cite{RuggeriBlake2020}.

Section  2 lists the evolution equation of $\xi(r,t)$
in the Gaussian approximation.
Section 3 presents  the twin-peak initial power spectrum,
and specifies the ranges of the parameters.
Section 4 gives the solution
and compares with the observational data.
Section 5  gives the conclusion  and discussion.

\section{ The linear equation of two-point
   correlation function of galaxies }\label{sec2}

The system of galaxies in the expanding Universe can be modeled by
a Newtonian self-gravity fluid  \cite{LandauLifshitz,Peebles1980},
and the nonlinear field equation of the mass density is known \cite{Peebles1980}.
In our previous work \cite{2021ZhangLi},
applying the functional derivative
\cite{BinneyDowrickFisherNewman1992,Goldenfeld1992}
to the ensemble average of the field equation of the mass density,
we have derived the nonlinear equation
of the two-point connected  correlation function $\xi$ in the expanding Universe,
which is a hyperbolic,  differential-integro equation,
and contains nonlinear terms in integrations,
such as the three-point  and four-point correlation functions,
$G^{(3)}$ and $G^{(4)}$,  hierarchically.
The  nonlinear equation is complex and its solution will involve much computation.
The nonlinear terms will enhance the clustering
and affect the correlation function only at small scales,
as the static nonlinear solution indicates \cite{ZhangMiao2009,ZhangChen2015}.
Dropping the nonlinear terms,
the evolution equation in the Gaussian approximation is simple
and valid at large scales $r \gtrsim 10$Mpc where $\xi \ll 1$.
In this paper we work with the evolution equation of $\xi$
in the Gaussian approximation  \cite{2021ZhangLi}
\bl \label{linapprgtcs0}
\ddot \xi ({\bf x},t)
    +2  H \dot  \xi ({\bf x},t)
    - \frac{ c_{s0}^2}{a^{2+2\eta   }(t)}  \nabla^2  \xi ({\bf x},t)
    - \frac{4 \pi G     \gamma  \rho_c \Omega_m }{a^{3}(t)} \,   \xi ({\bf x},t)
&  =  \frac{4 \pi G m}{ a^3 } \delta^{(3)}({\bf x}) ,
\el
where  $H=\dot a/a$,  $\rho_c$ is the critical density,
$m$ is the mass of typical galaxy,
$\gamma$ is the overdensity parameter ($\gamma \geq 1$),
$c_{s0}$ is the present sound speed of the system of galaxies,
and $\eta =1, \frac35$ for galaxies without or with self-rotation respectively.
(See the details in  Ref.\cite{2021ZhangLi}.)
The scale factor $a(t)$ is given by
$ \big(\frac{\dot{a}}{a} \big)^2
  = \frac{8\pi G}{3} \rho_c [a^{-3}\Omega_m +\Omega_\Lambda]$
with  $\Omega_m +\Omega_\Lambda=1$ and  $\Omega_\Lambda \simeq 0.7$.
The normalization $a=1/(1+z)$ will be used.

The evolution equation \eqref{linapprgtcs0} is a linear, hyperbolic equation,
and holds for the system of galaxies  on large scales
inside the horizon of the expanding Universe.
The term $2 H \dot \xi$ is due to the cosmic expansion,
and has the  effect of suppressing the growth of clustering.
The pressure term $- c_{s0}^2  \nabla^2 \xi$
gives rise to acoustic oscillations on the large scales,
and its role is against the clustering.
The gravity term $- 4 \pi G  \gamma \rho_0 \xi$
is  the main driving force for clustering.
The inhomogeneous term $4 \pi G m \delta^{(3)}({\bf x})$
is a source for the correlation function,
and enhances the amplitude of correlation function \cite{2021ZhangLi}.
Its  magnitude is proportional to $m$,
and explains the observed fact that galaxies of higher mass
acquire a higher correlation amplitude
\cite{HauserPeebles1973,KlypinKopylov1983,BahcalSoneira1983,Bahcal1996}.
In the static case $\ddot \xi = \dot \xi =0$,
Eq.\eqref{linapprgtcs0} will reduce to the static linear equation
that was studied in Ref.\cite{Zhang2007}.
By comparison,   in the linear approximation of
Davis-Peebles' equation (72),
the pressure term was ignored as they considered a pressureless gas,
and the $\delta^{(3)}({\bf x})$ term was dropped in a massless limit.

The equation \eqref{linapprgtcs0} can be solved  conveniently in the $k$-space
without specifying the boundary condition.
Introduce the Fourier transformation,
\bl\label{Fouriertrafm}
\xi ({\bf x},t)=\frac{1}{(2\pi)^{3 }}
   \int d^3k   \,   P_k(t) e^{i \,\bf{k}\cdot\bf{x}}  ,
\el
where $P_k(t)$  is the power spectrum of dimension $[L^3]$,
which is related to  $\Delta^2_k \equiv \frac{k^3}{2\pi^2}  P_k $
often used in the literature.
Using  $a$ as the time variable,
Eq.\eqref{linapprgtcs0} leads to
the ordinary differential equation of the power spectrum
\bl
& \frac{\partial^{2}}{\partial a^{2}}  P_k
+ \Big( \frac{3}{a}-\frac{3}{2 a} \frac{a^{-3}
       \Omega_{m}}{ (a^{-3} \Omega_{m}+\Omega_{\Lambda} )} \Big)
           \frac{\partial}{\partial a} P_k
+  \frac{3}{2 a^{4+2\eta }} \frac{  \Omega_{m}}
     {(a^{-3}  \Omega_{m}+\Omega_{\Lambda})}
       \Big( \frac{k^2}{k_{J}^2}  - \gamma\,  a^{2\eta -1}   \Big) P_k
       \nn \\
& ~~~~~  =\frac{A}{a^{5} (a^{-3} \Omega_{m}+\Omega_{\Lambda} )} ,
\label{linequavar}
\el
where $A \equiv 4\pi Gm/ H_0^2$ with  $H_0$ being the Hubble constant,
\be \label{kJdef}
k_{J} \equiv  \frac{\sqrt{ 4 \pi G  \,    \rho_c \Omega_m}}{ c_{s0} }
= \big(\frac32 \Omega_m \big)^{1/2} \frac{H_0}{c_{s0}}
\ee
is the present Jeans wavenumber,
and $\lambda_J= 2\pi/k_J$ is the present  Jeans length
of the system of galaxies.
In the expanding Universe,  the Jeans length  is stretching as
$\lambda_J(t)=  a^{\frac32-\eta}  \lambda_J $,
 slower than the comoving ($\propto a$).
For each  $k$, Eq.\eqref{linequavar} describes an oscillating mode
when $k^2/k^2_{J} > \gamma \,  a^{2\eta -1}$,
or a growing mode when  $k^2/k^2_{J} < \gamma \,  a^{2\eta -1}$.
Asides the cosmological parameters  $H_0$ and $\Omega_m$,
Eq.\eqref{linequavar} contains
$m$,  $k_J$,  $\gamma$ and $\eta$ as the parameters of
the self-gravity system of galaxies.

\section{ The initial condition and the parameters }\label{sec3}

An appropriate  initial power spectrum  of galaxies is needed
to solve the differential equation \eqref{linequavar}.
Currently we do not know
the observed correlation function of galaxies
at the early stage of formation \cite{Peebles1993}.
Observations indicate that
galaxies are formed  not later than
a redshift range $z \sim 10-13$ \cite{Bunker:2023lzn, Robertson:2022gdk}.
In our previous work  \cite{2021ZhangLi}
a simple one-peak initial power spectrum  of galaxies was used at $z=7$,
which corresponds to the first characteristic peak of
the primeval spectrum of BAO  at the decoupling  ($z\sim 1000$)
\cite{SunyaevZeldovich1970,PeeblesYu1970}
that have survived the Silk damping \cite{Silk1968,Field1971,Weinberg1971}.
In fact, the primeval spectrum of BAO contains  multi characteristic  peaks.
For instance,  Ref.\cite{PeeblesYu1970} in their Figure 4
gave four BAO peaks which survive the Silk damping,
for a model of baryon $\Omega_b=0.03$,
with the first two peaks corresponding to the comoving wavelengths
$100 h^{-1}$Mpc, $140 h^{-1}$Mpc approximately,
(other two peaks corresponding to
the wavelengths which are too long and beyond
the range of the current galaxy surveys.)
The details of these BAO peaks are model-dependent,
and sensitive to the fractions of the baryon,   etc.
For a CDM model with $\Omega_b= 0.1, \Omega_c=0.1$,
Ref.\cite{EisensteinHu1998} gave a first peak at $78 h^{-1}$Mpc
and a broad second peak at  $\gtrsim 150  h^{-1}$Mpc.
(See also Refs.\cite{HuSugiyama1995,Holtzman1989} for other model parameters.)
It is expected that, after the decoupling,
 the BAO imprints  are eventually transferred to
the initial power spectrum  of the system of galaxies.
(The detail of the transferring process is worthy of further study.)
The point is that two peaks of the primeval BAO spectrum are relevant,
as they fall into the range of the current galaxy surveys.

\begin{figure}
    \centering
    \includegraphics[width=0.45\textwidth]{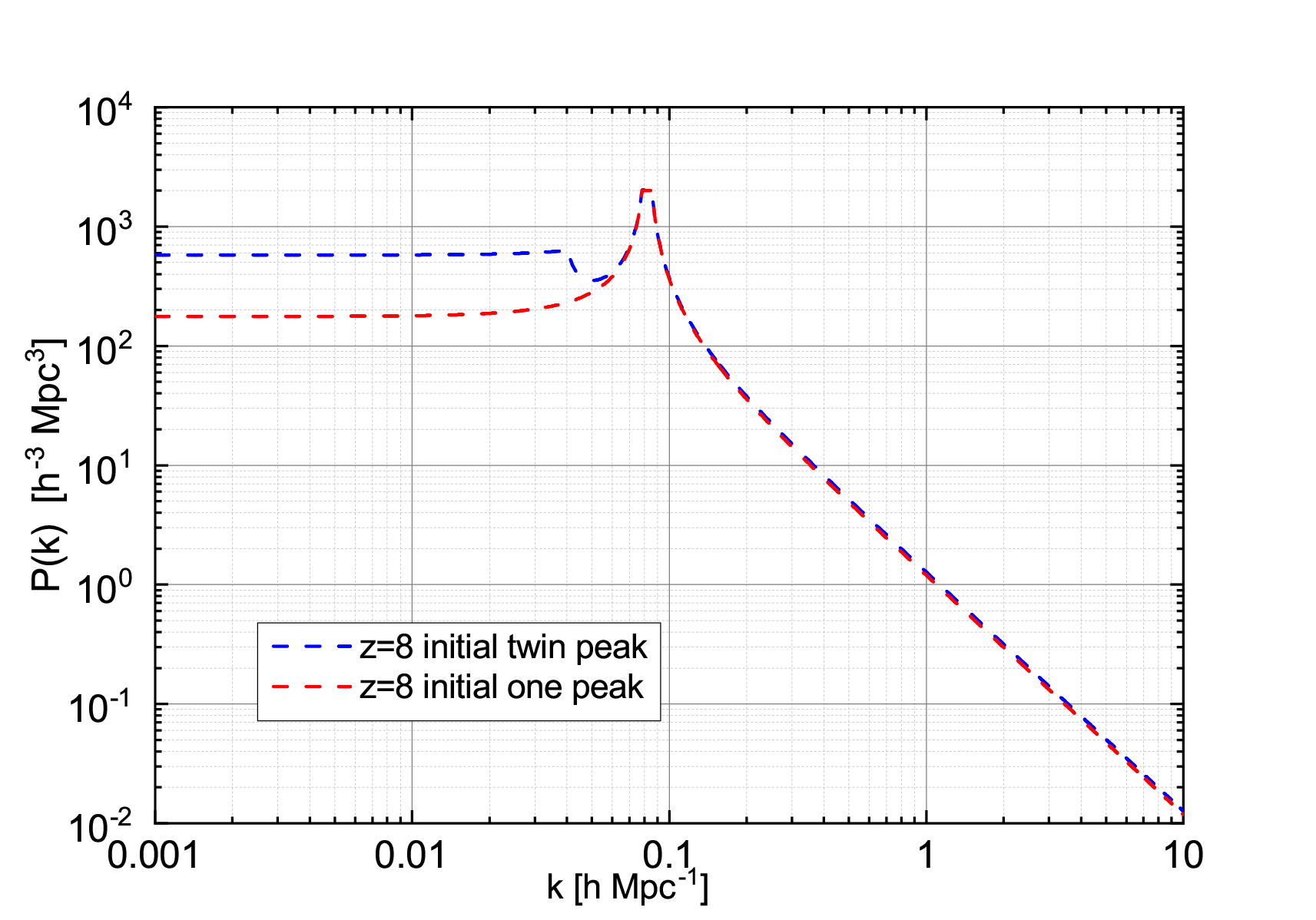}
    \caption{Blue dashed line:
    the twin-peak initial spectrum  $P_{k\, ini}$ at $z=8$
 adopted in this paper.
      Red dashed line: the one-peak initial spectrum used in Ref.\cite{2021ZhangLi}.
    }
    \label{fig:ini}
\end{figure}

Motivated by  the above analysis,
we adopt a twin-peak initial power spectrum of
the system of galaxies as the following
\bl \label{P_i2}
& P_{k\,  ini}(z)=\frac{1}{2 (1+z)^3 n_0}
\Big[\frac{1}{ \big|\left(\frac{k}{k_1}\right)^2- 1 \big|}
+\frac{\epsilon(k)}{\big|\left(\frac{k}{k_2}\right)^2- 1 \big|}\Big] ~.
\el
The first term in \eqref{P_i2} is the one peak initial spectrum
that was used in Ref.\cite{2021ZhangLi},
and the second term gives
a second  broad peak which is newly added in this paper,
 $n_0$ is the present number density of galaxies
(we take  $1/2 n_0  \sim  5 \times 10^4 \,  h^{-3}$Mpc$^{3}$),
$k_1$ and $k_2$ are the locations of the first and second peaks, and
$\epsilon (k)$ is a step-like function to give a broad second peak,
as in the primeval BAO spectrum in Refs.\cite{EisensteinHu1998,HuSugiyama1995,Holtzman1989}.
We shall choose  an initial  redshift $z=8$ for actual computing in this paper,
and  take $k_1$ and $k_2$ as parameters to fit with
the observational data \cite{RuggeriBlake2020}.
Approximately,  the first peak at $k_1 \sim  0.083 h \mathrm{Mpc}^{-1}$
 corresponds a comoving wavelength $\sim 75\,  h^{-1}$Mpc,
and the second peak at $k_2 \sim  (0.026\sim 0.030)  h \mathrm{Mpc}^{-1}$
corresponds to a  comoving wavelength $(210\sim 250) h^{-1}$Mpc.
The concrete expression  \eqref{P_i2} is based on
the analytic static solution in Ref.\cite{Zhang2007},
and has a similar profile to the initial linear spectrum
used in simulations \cite{Springel2018}.
The form of the expression \eqref{P_i2} is not unique,
and other possible choices are allowed.
The divergences at $k=k_1$ and $ k=k_2$
in the expression \eqref{P_i2} will be cut off in computing.
Fig.~\ref{fig:ini} shows the initial twin-peak spectrum
$P_{k\,  ini}$ at $z=8$ in the blue solid line,
and the red dashed line is the one peak spectrum used in Ref.\cite{2021ZhangLi}.

To solve  \eqref{linequavar}, an initial rate is also needed.
The rate $r_a$ is defined by
\be
\frac{\partial}{\partial a}   P_k (a) =  r_a   P_k (a)  ,
\label{ratedef}
\ee
The rate is affected by the cosmological constant $\Omega_\Lambda$ \cite{Lahav1991}.
For $\Omega_\Lambda\simeq 0.70$ and $0.75$,
the initial rate at $z=8$ is $r_a \simeq 0.48$ and $0.52$, respectively.
Note that $r_a \simeq 0.50$ at $z=7$,
which corrects a typo in Ref.\cite{2021ZhangLi}.
As it turns out, the outcome at low redshifts
is not very sensitive to the value of initial rate.

In our computing, the   cosmological parameters are taken
in the range  $\Omega_m = (0.25 \sim 0.30)$,
and $H_0^{-1} =3000 \, h^{-1}$ Mpc with $h = (0.69\sim 0.73)$.
To fit with the observational data of
the correlation function  from the WiggleZ surveys \cite{RuggeriBlake2020},
the fluid parameters  can be taken in the range
$k_J = ( 0.045  \sim 0.090 ) \,  h\, \mathrm{Mpc}^{-1}$,
i.e., $\lambda_J= (139 \sim 71) \,  h^{-1} \text{Mpc} $,
$A = (2 \sim 4) \times 10^3 \,  h^{-2} \mathrm{Mpc}^3 $,
$\gamma = 1 \sim 6 $.
These ranges of parameters are similar to
those used in the previous work \cite{2021ZhangLi}.

\begin{figure}
    \centering
    \includegraphics[width=0.47\textwidth]{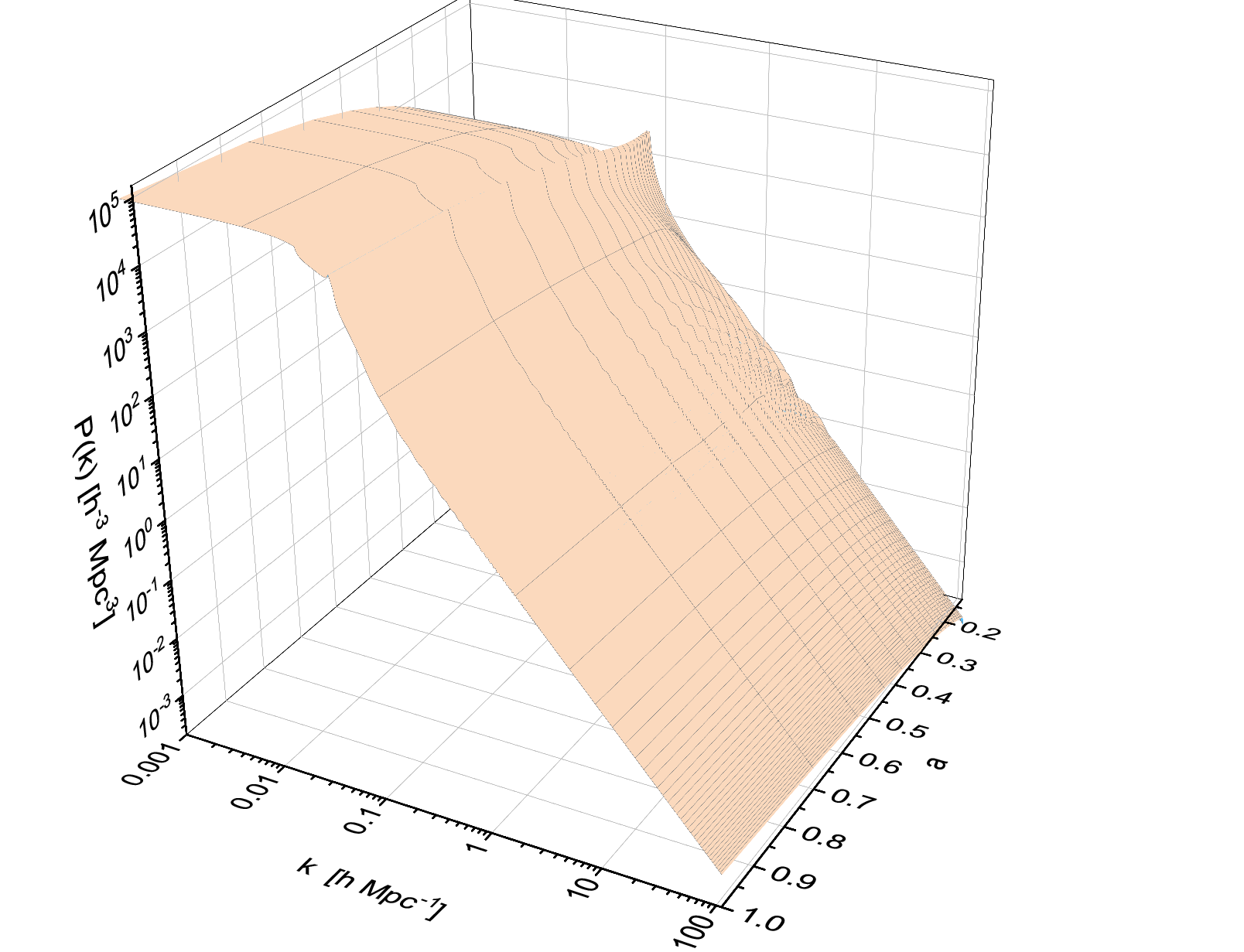}
    \hspace{0.2in}
    \includegraphics[width=0.48\textwidth]{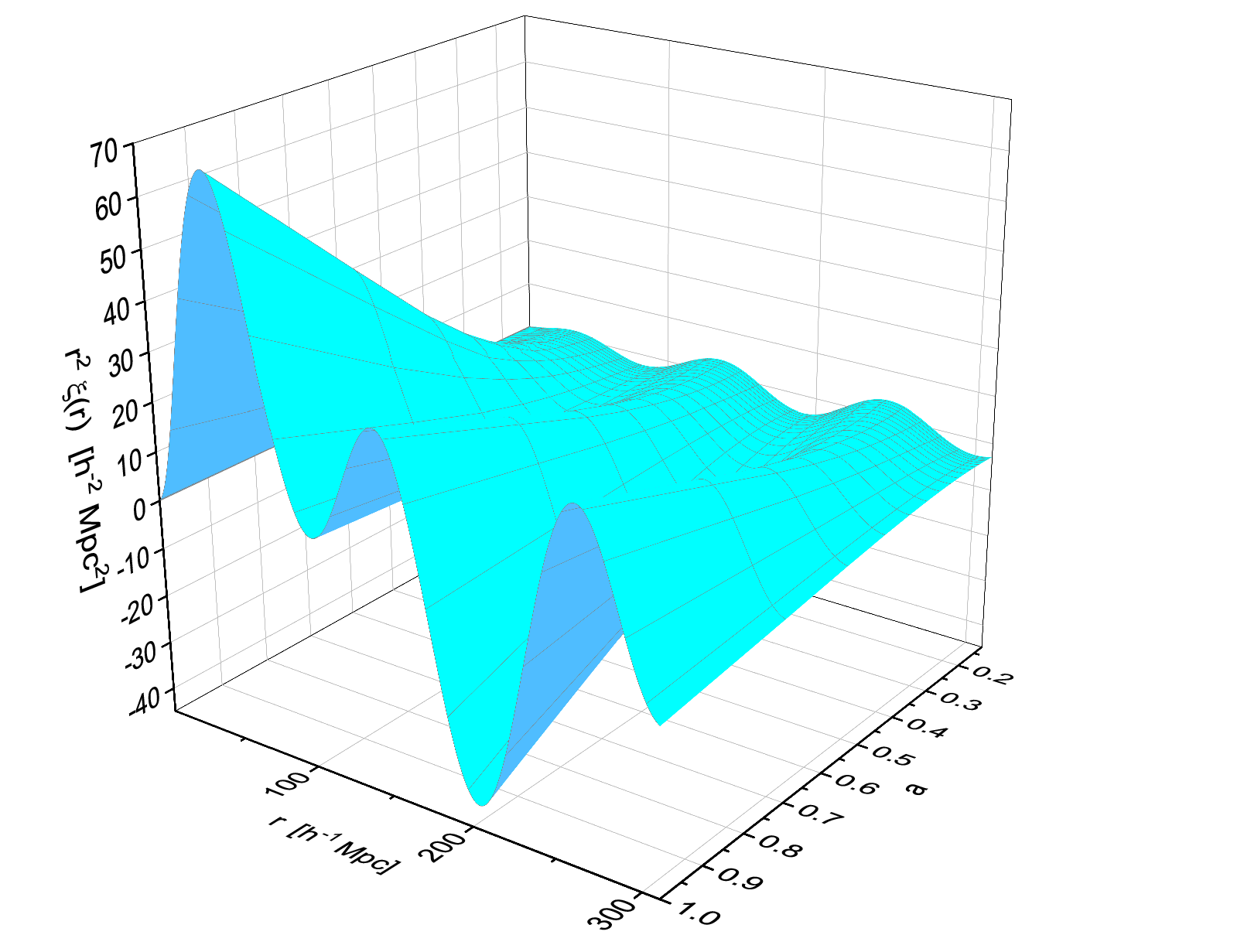}
    \caption{ The whole  evolution (from $z=8 $ to $z=0$)
      for the model  $c_s \propto  a^{-1}$.
     Left panel: $P_k(a)$.  Right panel:  $r^2\xi(r,a)$.
      }
    \label{3Dcs1}
\end{figure}

\begin{figure}
    \centering
    \includegraphics[width=0.47\textwidth]{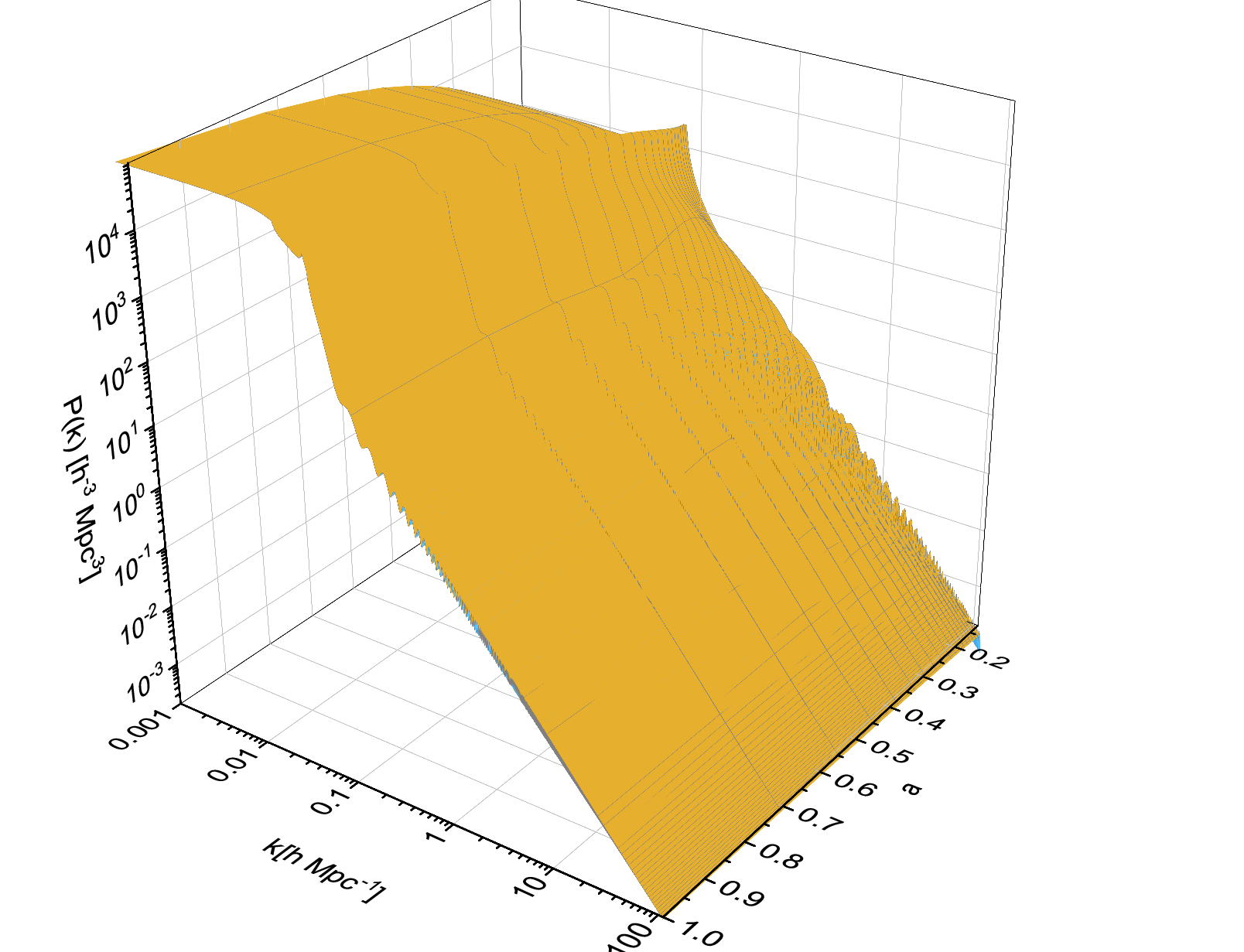}
    \hspace{0.2in}
    \includegraphics[width=0.48\textwidth]{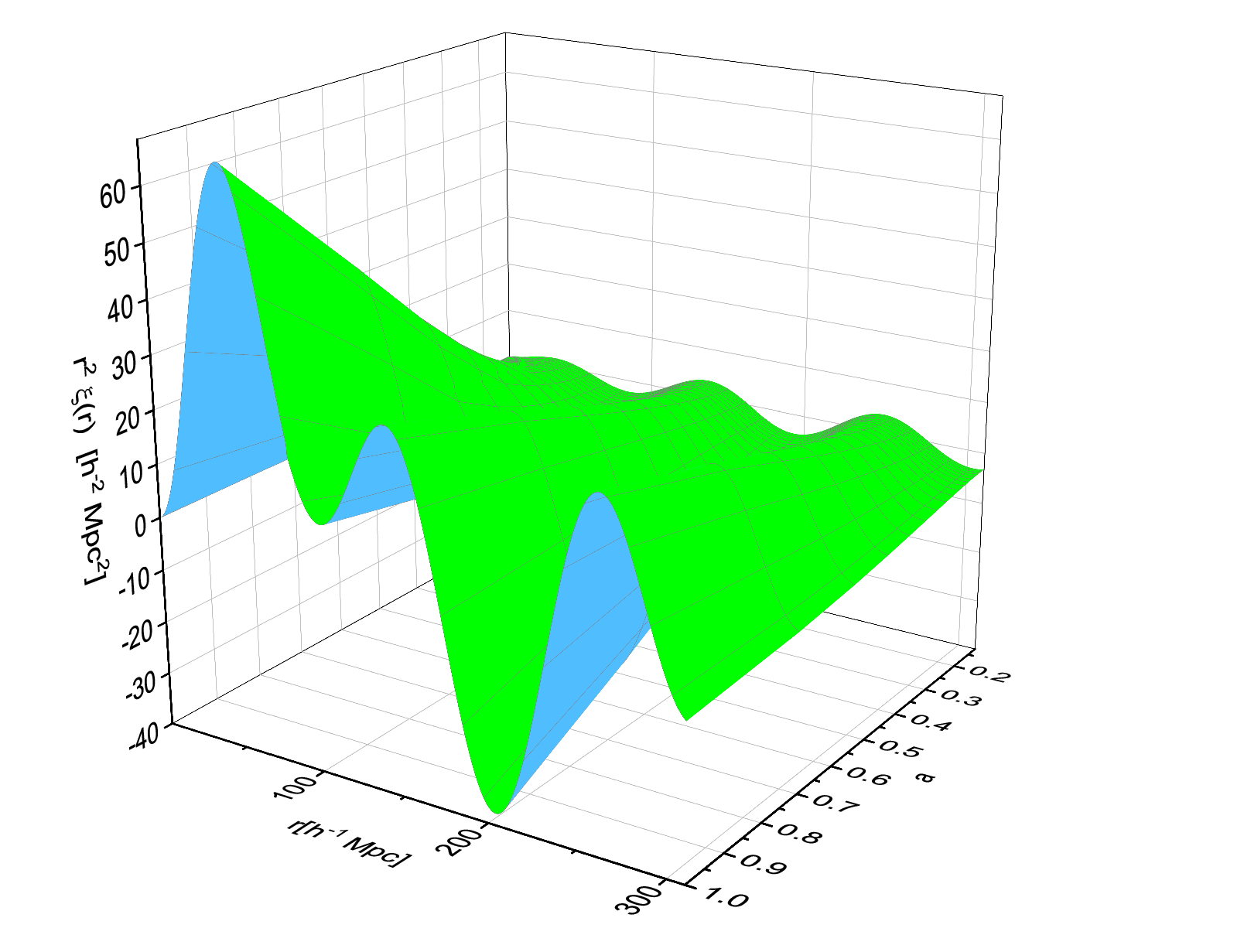}
    \caption{ The whole evolution  (from $z=8 $ to $z=0$)
    for the model  $c_s \propto a^{-3/5}$.
     Left panel: $P_k(a)$. Right panel:  $r^2\xi(r,a)$.
     }
    \label{3Dcs2}
\end{figure}

\section{The Results}\label{sec5}

Given the equation \eqref{linequavar},
together with the twin-peak initial condition \eqref{P_i2} \eqref{ratedef} at $z=8$,
we obtain numerically the solution $P_k(a)$ for each $k$,
and  $\xi(r,a)$ by Fourier transformation.
The whole evolution from $z=8$ to $z=0$
of the spectrum $P_k(a)$ and the weighted  $r^2 \xi(r, a)$ are shown
in Fig.\ref{3Dcs1}  for the model  $c_s\propto a^{-1}$,
and in Fig.\ref{3Dcs2} for the model $c_s\propto a^{-3/5}$,  respectively.
During the evolution, $P_k(a)$ keeps a profile
similar to the initial $P_{k\, ini}$, and is increasing in amplitude.
 $\xi(r,a)$ contains periodic bumps and troughs,
the bumps  are  getting higher and the troughs are getting deeper,
and the separation between bumps is stretching to a greater distance.
The evolution is quite smooth and there are no abrupt jumps,
and in this sense the distribution of galaxies is in an asymptotically relaxed state
 in the expanding Universe  \cite{Saslaw1985}.

\begin{figure}
    \centering
    \includegraphics[width=0.45\textwidth]{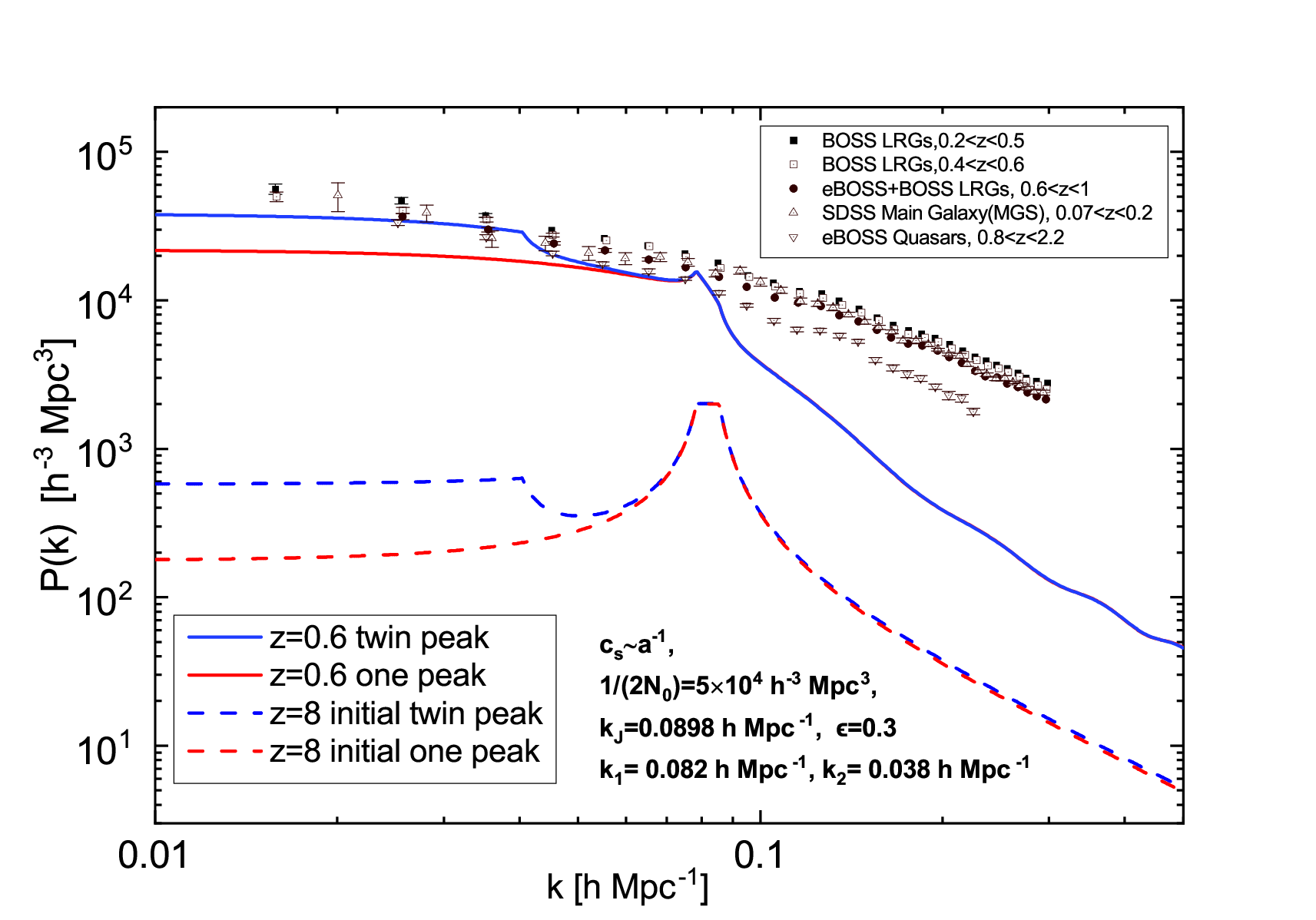}
    \hspace{0.2in}
    \includegraphics[width=0.45\textwidth]{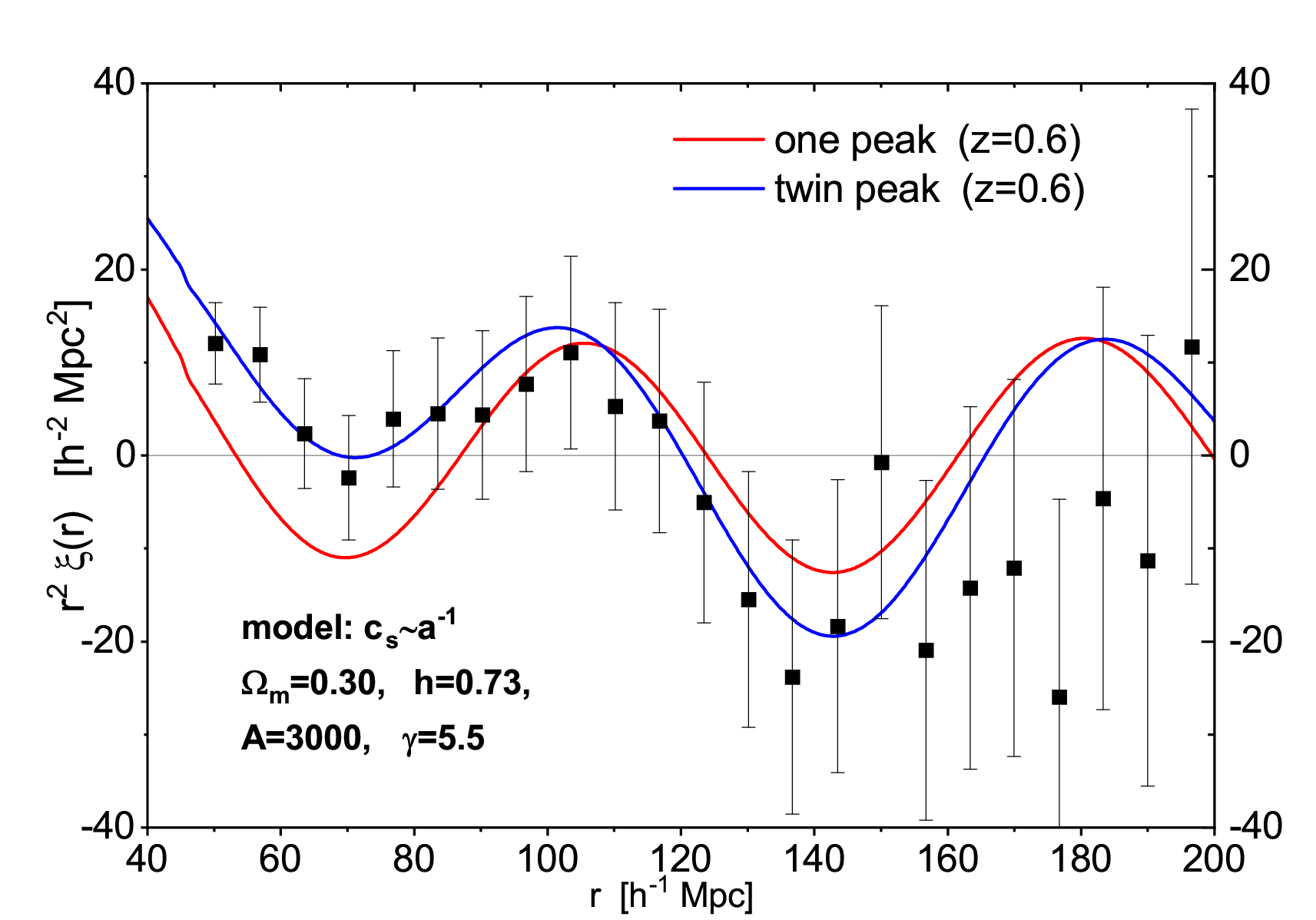}
    \caption{ The model $c_s \propto a^{-1}$.
Left panel: $P_k$ and the data from
    Refs.\cite{Alam2017,Ross2015,Neveux2020,Bautista2021,Gil-Marin2020}.
Right panel: $r^2 \xi(r)$ and the data from WiggleZ \cite{RuggeriBlake2020}.
The output is taken at a redshift $z=0.6$
    to compare with the observational data  \cite{RuggeriBlake2020}.
The parameters:  $k_J = 0.0898 h \mathrm{Mpc}^{-1}$,
       $k_1 = 0.082 h  \mathrm{Mpc}^{-1}$, $k_2 = 0.038 h  \mathrm{Mpc}^{-1}$,
       $1 / (2 n_0) = 5 \times 10^4 h^{-3} \mathrm{Mpc}^3$, $\epsilon =0.3 $.
     }
    \label{fig:cs1}
\end{figure}

\begin{figure}
    \centering
    \includegraphics[width=0.45\textwidth]{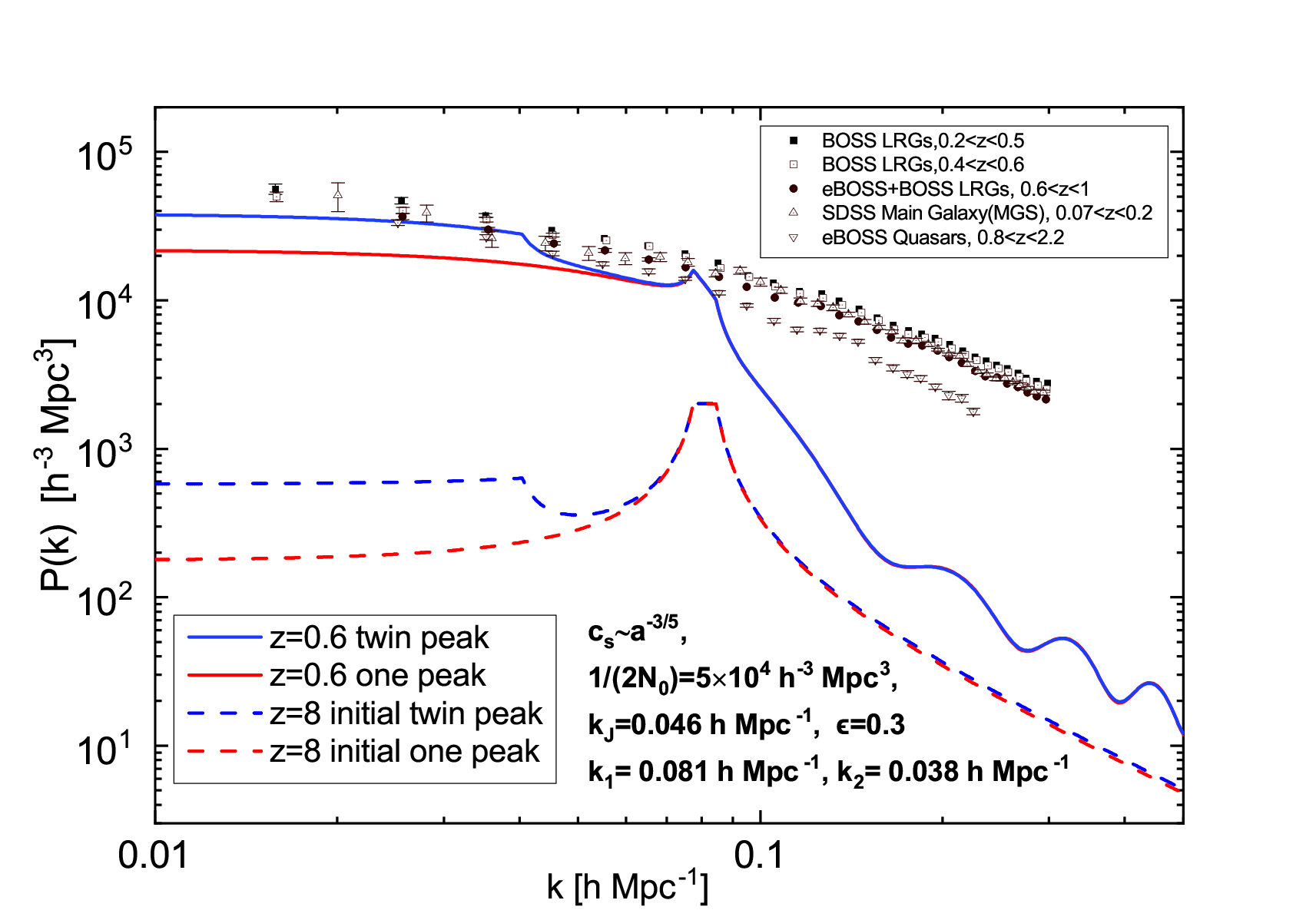}
    \hspace{0.2in}
    \includegraphics[width=0.45\textwidth]{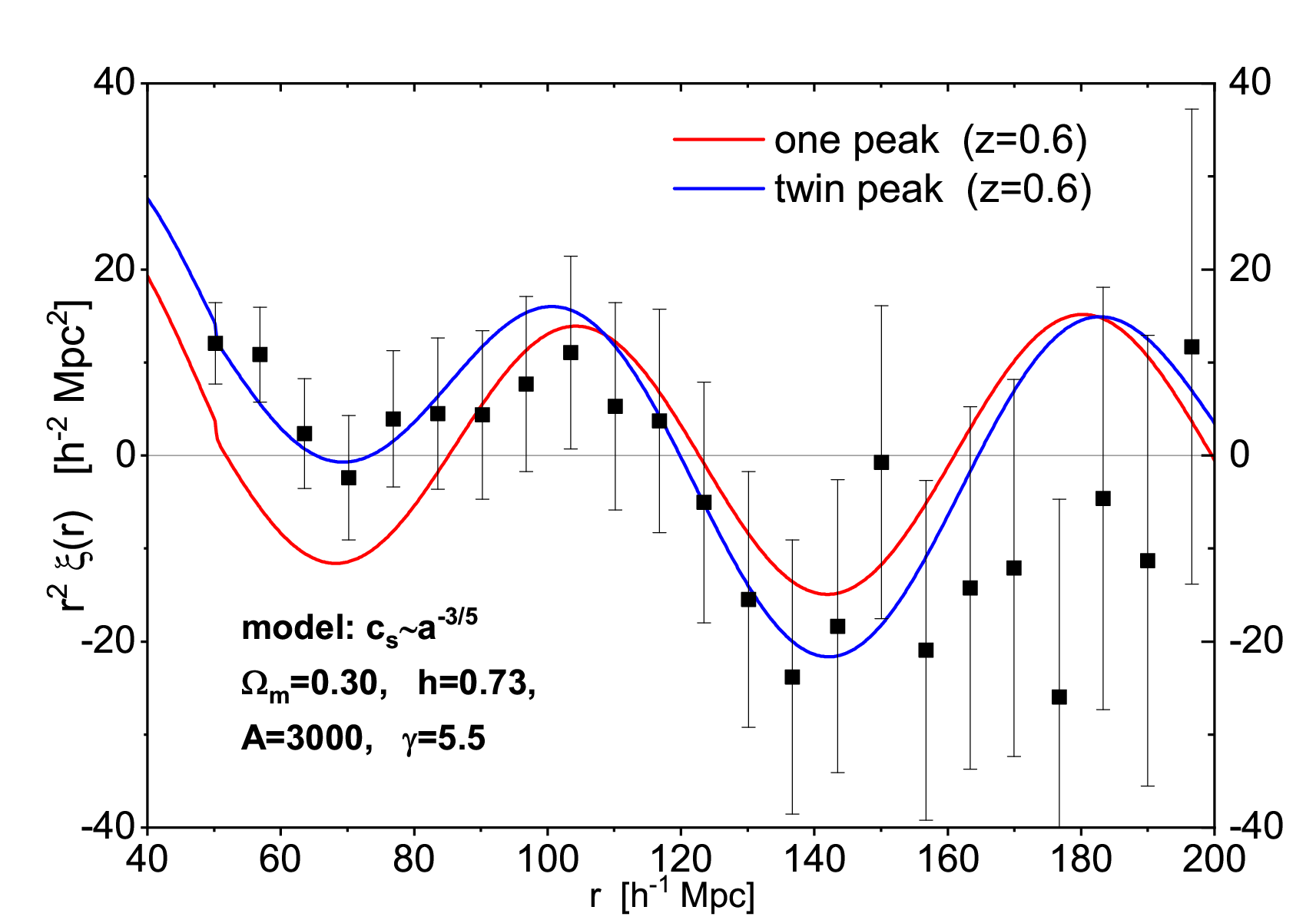}
    \caption{ The  model $c_s \propto a^{-3/5}$.
    Left panel: $P_k$ and the data from
    Refs.\cite{Alam2017,Ross2015,Neveux2020,Bautista2021,Gil-Marin2020}.
    Right panel: $r^2 \xi(r)$ and
    the data from WiggleZ \cite{RuggeriBlake2020}.
     The output at $z=0.6$.
The parameters:  $k_J = 0.046 h \mathrm{Mpc}^{-1}$,
       $k_1 = 0.081 h  \mathrm{Mpc}^{-1}$, $k_2 = 0.038 h  \mathrm{Mpc}^{-1}$.
        }
    \label{fig:cs35}
\end{figure}

To compare with the WiggleZ data with a mean redshift $z\sim 0.6$ \cite{RuggeriBlake2020},
we plot the output at $z=0.6$ in Fig.\ref{fig:cs1}
for  $c_s\propto a^{-1}$
and in Fig.\ref{fig:cs35} for $c_s\propto a^{-3/5}$,
and the outcome by the one-peak initial spectrum  \cite{2021ZhangLi}
is also given for comparison.
Prominently,
 the $100 \mathrm{Mpc}$ periodic bumps of the weighted $r^2\xi(r)$
occur and are located  approximately at
\bl
r \sim    100,  \, 200 ,   \, 300 \,  h^{-1}   \text{Mpc},
\label{bumpsloc}
\el
the bump separation $\Delta r \sim 100 h^{-1}$ Mpc
which is identified as the Jeans length $\lambda_J$ in eq.\eqref{linequavar}.
This $100 \mathrm{Mpc}$ periodicity
occurs also in the static solution \cite{Zhang2007,ZhangMiao2009,ZhangChen2015},
and in the evolution solution with the one peak initial spectrum \cite{2021ZhangLi},
and now shows up the solution with the twin-peak initial spectrum.
Early pencil-beam redshift surveys already observed the 100Mpc periodic feature
in the correlation function of galaxies
 \cite{Broadhurst1990,Broadhurst1995,Tucker1997}
 and of clusters  \cite{Einasto1997b,Einasto2002b,Tago2002}.
Recent surveys observe two bumps,
one  at $\sim 100 h^{-1}$Mpc and another at $\sim 200 h^{-1}$ Mpc,
in the correlation function of galaxy
\cite{RuggeriBlake2020,Beutler6dFGS2011,Anderson2014,Anderson2012,SanchezSDSS2017},
as well as of quasars
\cite{Ata2018,BausticaBusca2017,Busca2013,Slosar2013,Agathe2019}.
All these observational  results  confirm
the predicted 100Mpc periodic feature  \eqref{bumpsloc}.
Some simulations also show this phenomenon \cite{Yahata2005}.
In the future there may be a chance of detecting
the third bump  at $\sim 300 h^{-1}$ Mpc
as our result  \eqref{bumpsloc} predicts.
Our computing shows that
the  100Mpc periodicity
is also slightly modified by the initial spectrum and by other parameters as well.

An important result by using the twin-peak initial spectrum
is the prediction of a shallow trough at  $ \sim 70 h^{-1}$Mpc
and a deep trough at $\sim 140 h^{-1}$Mpc in the correlation function.
This results from  the superposition of the twin peaks of the initial spectrum.
The second peak at $k_2$ corresponds
to a very long wavelength $(210\sim 250) h^{-1}$Mpc,
affects  the profile due to the first peak,
and consequently modifies  the depths of the two troughs
at $\sim 70 h^{-1}$Mpc  and $\sim 140 h^{-1}$Mpc.
The right panels of Fig.\ref{fig:cs1} and Fig.\ref{fig:cs35} show that
the profiles  of the two troughs
match the data of WiggleZ \cite{RuggeriBlake2020},
much better than those from the one-peak initial spectrum \cite{2021ZhangLi}.
The outcomes from the two initial power spectra
can be compared by  using the chi-square as a statistical test,
\bl
\chi^2 =\sum_{i=1}^N \frac{(x_{o,  i} - x_{t, i})^2}{\sigma^2_i},
\el
where $x_{o, i}$ is the observed value,
$x_{t, i}$ is the theoretical value,
and $\sigma^2_i$ is the variance of observed value.
In the model  $c_s \propto a^{-1}$,
we get $\chi^2=8.8$ for the twin-peak initial spectrum,
and  $\chi^2=30.8$ for the one-peak initial spectrum,
and respectively, in the  model $c_s \propto a^{-3/5}$,
 $\chi^2=11.3$ for the twin-peak initial spectrum,
and  $\chi^2=34.4$ for the one-peak initial spectrum.
So, statistically
 the twin-peak initial spectrum gives a better account of
the observational data
than the one-peak initial spectrum.

Although we have chosen the parameters
to fit the observed $\xi(r)$ of the WiggleZ \cite{RuggeriBlake2020},
nevertheless, the associated outcome $P_k$  at large scales
is also qualitatively consistent with the observed power spectrum
of  eBOSS and SDSS  \cite{Alam2017,Ross2015,Neveux2020,Bautista2021,Gil-Marin2020}.
At small scales ($k \geq 0.1 h$Mpc$^{-1}$)
$P_k$ is lower than the data,
and this is the limitation of the Gaussian approximate eq.\eqref{linequavar}
in which the nonlinear terms have not been included  \cite{2021ZhangLi}.

\section{Conclusion and discussion}\label{sec6}

We study the evolution equation of
the correlation function of galaxies in the Gaussian  approximation,
and use the twin-peak initial power spectrum  \eqref{P_i2} at $z=8$,
which inherits a relevant portion of the imprints of
the primeval BAO spectrum \cite{PeeblesYu1970,EisensteinHu1998}.
The evolution of the power spectrum  $P_k(a)$
and the  two-point correlation function $\xi(r,a)$ are obtained.
 $\xi(r)$  not only contains  the $100$ Mpc periodic  bumps,
but also predicts a shallow trough at $\sim 70 h^{-1}$Mpc
and a deep trough at $\sim 140 h^{-1}$Mpc,
agreeing  with the observational data of  WiggleZ \cite{RuggeriBlake2020}
 up to the range  $\lesssim  200 h^{-1}$Mpc.
The outcome  improves substantially the previous work that used
the simple one-peak initial spectrum \cite{2021ZhangLi}.

The periodic bump separation $\Delta r \sim 100 h^{-1}$ Mpc
is determined by the Jeans wavelength $\lambda_J$ in Eq.\eqref{linequavar},
mildly modified by the first peak at $k_1$ (short-wavelength)
of the initial power spectrum and the parameters as well.
This explains why the previous work \cite{2021ZhangLi}
with the one-peak  initial spectrum also showed the 100 Mpc periodic feature.
The second peak at $k_2$ of the twin-peak initial spectrum
corresponds to a much longer wavelength $(210\sim 250) h^{-1}$Mpc,
and its associated  waves will superimpose upon
the 100 Mpc periodic bumps,
so that  the  depths of the two troughs at
$70 h^{-1}$MPc, $140 h^{-1}$Mpc are corrected,
as are observed by the WiggleZ.
A great advantage of the analytical equation of the correlation function
is that the solution  $\xi(r)$
explains the observational data in a simple, direct manner.

Since the discovery of the $100$ Mpc periodic feature
in 1990s \cite{Broadhurst1990,Broadhurst1995},
it has been interpreted by various tentative models
 \cite{Tucker1997,Einasto1997b,Einasto2002b,Tago2002}.
More recently it was interpreted as
the imprint of the sound horizon at the decoupling
\cite{Eisenstein2005,DHWeinberg2013},
which is the comoving distance $s$
that baryon acoustic sound waves travel.
However, there are two obvious difficulties
with the sound horizon interpretation.
First, the wave in the baryon-photon plasma
is stochastic and its path is unobservable statistically,
as is the distance of the path.
Thus, the sound horizon by definition is unobservable.
What can be possibly measured is the characteristic peaks
of the primeval spectrum of BAO \cite{SunyaevZeldovich1970}.
The twin-peak initial spectrum that we use
incorporates relevant imprints of the primeval BAO spectrum.
Second,
the sound horizon is predicted to be $s \simeq 167$ Mpc \cite{DHWeinberg2013},
which is much higher than the observed $100$ Mpc feature
in the correlation function of galaxies.
Moreover, the sound horizon
can not  explain the observed troughs
at $70 h^{-1}$Mpc and $140 h^{-1}$Mpc,
nor the observed second bump at $\sim 200 h^{-1}$Mpc.

Although the solution explains the observed features on large scales,
but, on small scales,
$P_k$ is lower than the observed spectrum.
We expect that the small-scale behavior  will be improved
by the nonlinear evolution equation    \cite{2021ZhangLi},
as has been the case in the static nonlinear solution \cite{ZhangMiao2009,ZhangChen2015}.

\section*{Acknowledgements}

Y. Zhang is supported by National Natural Science Foundation of China,
Grants No. 11675165,  No.  11961131007, and No.12261131497,
and in part by National Key RD Program of China (2021YFC2203100).

\end{document}